\documentclass{article}

\usepackage{PRIMEarxiv}

\usepackage[utf8]{inputenc}
\usepackage[T1]{fontenc}
\usepackage{amsmath}
\usepackage{amssymb,amsfonts}
\usepackage{mathrsfs}
\usepackage{hyperref}
\usepackage{url}
\usepackage{booktabs}
\usepackage{nicefrac}
\usepackage{microtype}
\usepackage{lipsum}
\usepackage{fancyhdr}
\usepackage{graphicx}
\usepackage{orcidlink}
\usepackage{multirow}
\usepackage[title]{appendix}
\usepackage{xcolor}
\usepackage{textcomp}
\usepackage{manyfoot}
\usepackage{algorithm}
\usepackage{algorithmicx}
\usepackage{algpseudocode}
\usepackage{listings}
\usepackage{tabularx}
\graphicspath{{media/}}     

\title{Emotional Vietnamese Speech-Based Depression Diagnosis Using Dynamic Attention Mechanism}

\author{
 Quang-Anh N.D.\orcidlink{0009-0003-1353-3367}, Manh-Hung Ha* \orcidlink{0000-0002-5782-6829}, Thai Kim Dinh \orcidlink{0000-0002-9060-4769}, Minh-Duc Pham, Ninh Nguyen Van \\
  International School, Vietnam National, University, Hanoi, Vietnam \\
  \texttt{anhnd@vnuis.edu.vn, hunghm@vnu.edu.vn*, thaikd@vnu.edu.vn,} \\
  \texttt{pmduc2808@gmail.com, ninhnv@vnuis.edu.vn}}

\begin{document}
\maketitle

\begin{abstract}
    Major depressive disorder is a prevalent and serious mental health condition that negatively impacts your emotions, thoughts, actions, and overall perception of the world. It is complicated to determine whether a person is depressed due to the symptoms of depression not apparent. However, their voice can be one of the factor from which we can acknowledge signs of depression. People who are depressed express discomfort, sadness and they may speak slowly, trembly, and lose emotion in their voices. In this study, we proposed the Dynamic Convolutional Block Attention Module (Dynamic-CBAM) to utilized with in an Attention-GRU Network to classify the emotions by analyzing the audio signal of humans. Based on the results, we can diagnose which patients are depressed or prone to depression then so that treatment and prevention can be started as soon as possible. The research delves into the intricate computational steps involved in implementing a Attention-GRU deep learning architecture. Through experimentation, the model has achieved an impressive recognition with Unweighted Accuracy (UA) rate of 0.87 and 0.86 Weighted Accuracy (WA) rate and F1 rate of 0.87 in the VNEMOS dataset. Training code is released in https://github.com/fiyud/Emotional-Vietnamese-Speech-Based-Depression-Diagnosis-Using-Dynamic-Attention-Mechanism
\end{abstract}


\section{Introduction}

    The word "Melancholy" was first used by the ancient Greeks to describe feeling intensely sad and despair, to be more familiar in modern times, this word can be called "Depress". The mental illness of depression, considered to be one of the most prevalent and serious mental illnesses, severely impairs an individual's quality of life, marking them with feelings of sadness, hopelessness, and loss of interest in enjoyable activities. The diagnosis of schizophrenia requires at least five or more of these symptoms occurring for at least two weeks: depressed mood, diminished interest pleasure, significant weight loss or gain, insomnia, hypersomnia, psychomotor agitation, fatigue, feelings of worthlessness, diminished concentration, and recurrent thoughts of suicide or death \cite{b1}. Various factors are involved in the origins and causes of depression, including genetic, biological, environmental, and psychological factors \cite{b2} With a lifetime prevalence of around 15-20\%, depression contributes significantly to the global burden of disease and affects individuals of all ethnicities, ages, and genders \cite{b3}. The World Health Organization reports that depression causes more than 300 million people around the world to suffer from disabilities, even though there are effective treatments available, less than half of those affected worldwide receive them \cite{b3}. Depression can be alleviated by improving access to evidence-based interventions. Clinical studies have demonstrated the efficacy of Cognitive Behavioral Therapy (CBT) and pharmacological treatments for treating acute depression episodes and preventing relapses, with CBT showing particular promise in maintaining long-term benefits \cite{b5}.

    Depressive disorders require profound compassion on the part of those suffering from their relentless and debilitating effects. A significant number of personal and societal costs will need to be alleviated through ongoing research and improved services. Although melancholy has perennially plagued humankind, hope persists in the progress made and the promise of better understanding and caring for those in its thrall. By using deep learning techniques, this research aims to assess the probability of individuals having a tendency to be depressed by analyzing the frequency amplitude of the voice. Our objective was to identify the speaker's specific emotional characteristics and then determine if they are likely to experience depression based on those characteristics. By using deep learning, we are able to assess and analyze the emotional characteristics of a voice in a more comprehensive and accurate manner. An evaluation of specific emotional characteristics in the human voice was performed, including speed, volume, rhythm, and tone. By the completion of this research, we will be able to obtain useful information to diagnose depression and be able to detect depression signs more accurately and quicker, further more Table ~\ref{tab1} outlined the number of people using antidepressants pill in Taiwan \cite{b6}.

\begin{table}[h!]
    \centering
    \caption{Statistics number of people using antidepressants pill in Taiwan.}
    \begin{tabular}{|l|l|c|c|c|c|}
        \hline
        \textbf{Country} & \textbf{2018} & \textbf{2017} & \textbf{2016} & \textbf{2015} & \textbf{2014} \\
        \hline
        Taiwan & 1,397,197 & 1,330,204 & 1,273,561 & 1,121,659 & 1,194,395 \\
        \hline
        Under 30 & 162,402 & 143,024 & 127,508 & 118,186 & 115,740 \\
        \hline
        \textbf{Ratio} & \textbf{11.62\%} & \textbf{10.57\%} & \textbf{10.01\%} & \textbf{9.75\%} & \textbf{9.69\%} \\
        \hline
    \end{tabular}
\label{tab1}
\end{table}

    Additionally, in 2014 E. Yüncü developed A binary decision tree consisting of SVM classifiers was utilized to classify seven emotions using the EmoDB database \cite{b7} reached highest accuracy of 82.9\%. Meanwhile, by utilizing multiple DNNs based on non-local attention and pretrained models, \cite{b7.1} achieves superior accuracy compared to \cite{b7}, with the highest recognition rate reaching approximately 93.2\%. In following years 2016 XingChan Ma used a neural network in conjunction with CNN and LSTM \cite{b8}, which achieved a resolution rate of 68\% for the DAIC-WOZ. In 2018, Haytham used IEMOCAP Database using CNN-RNN structure \cite{b9} and achieved a recognition rate of 64.78\%, in the same year, S. Tripathi used a three-layer LSTM architecture to perform emotional discrimination on IEMOCAP \cite{b10} and achieved a recognition rate of 71.04\%, followed by Toyoshima I et al. in 2023 proposed a speech emotion recognition model based on a multi-input deep neural network that simultaneously learns these two audio features by using CNN - DNN architecture and input is spectrogram and gemaps achived 61.49\% accuracy rate \cite{b11}, for better visualization, Table ~\ref{tab2} outlined the comparison of others method in recent years.

\begin{table}[h!]
    \centering
    \caption{Comparison of related works in recent years with abbreviations as follow anger (an), happiness (ha), boredom (br), disgust (dg), fear (fe), sadness (sd), neutral (nt), silence (sl).}
    \begin{tabular}{|l|l|l|l|l|l|}
    \hline
    \textbf{Ref} & \textbf{Emotion} & \textbf{Data} & \textbf{Language} & \textbf{Structure} & \textbf{Acc} \\
    \hline
    2014 & an, br, dg, fe, ha, sd, nt & EmoDB & German & Binary Tree & 82.9\% \\
    \hline
    2016 & ha, sd & DAIC-WOZ & English & CNN - LSTM & 68\% \\ 
    \hline
    2018 & an, ha, sd, nt & IEMOCAP & English & CNN - RNN & 64.78\% \\ 
    \hline
    2018 & an, ha, sd, nt & IEMOCAP & English & 3 Layers LSTM & 71.04\% \\
    \hline
    2023 & an, ha, sd, nt & IEMOCAP & English & CNN - DNN & 61.49\% \\
    \hline
    2024 & an, br, dg, fe, ha, sd, nt & EmoDB & German & multiDNNs+attention& 93.2\% \\
     \hline
    \end{tabular}
\label{tab2}
\end{table}
\section{Methodology}
\subsection{Dynamic Convolutional Block Attention Module}

\begin{figure}
    \centering
    \includegraphics[width=0.75\linewidth]{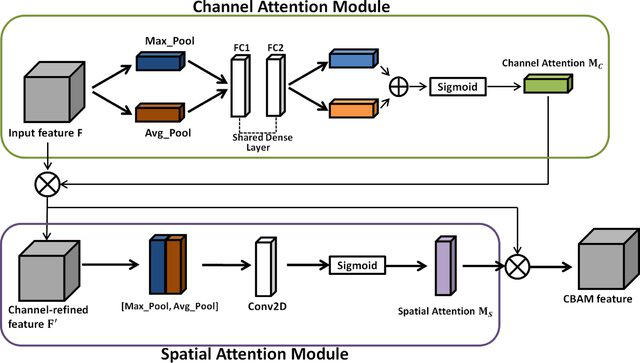}
    \caption{CBAM Block Structure \cite{b14}}
    \label{fig1}
\end{figure}

    In this paper, we proposed an new block new Dynamic-CBAM by modifying the Convolutional Block Attention Module (CBAM) \cite{b12} with Omni-Dimensional Dynamic Convolution (ODConv) \cite{b13}. The CBAM is an attention mechanism that enhances the feature of standard Convolutional Neural Networks (CNNs) by focusing on important features and suppressing unnecessary features. The CBAM block cotaining an hybrid attention mechanism by utilizing channel attention and spatial attention mechanism, given a feature map $F \epsilon R^{h \times w \times c}$ with $h, w, c$ denotes for height, weidth and channel, respectively and $f^{x \times x}$ stand for kernal size of the convolution layer. The channel attention and spatial attention modules in CBAM can be calculate by follow Equations.

\begin{equation}
\begin{aligned}
    M_c(F) & =\sigma\left(f^{1 \times 1}(\operatorname{Avg} \operatorname{Pool}(F))+f^{1 \times 1}(\operatorname{MaxPool}(F))\right) \\
    & =\sigma\left(W_1\left(W_0\left(F_{\text {avg }}^c\right)\right)+W_1\left(W_0\left(F_{\max }^c\right)\right)\right)
\label{eq1}
\end{aligned}
\end{equation}

\begin{equation}
\begin{aligned}
    M_s(F) & =\sigma\left(f^{7 \times 7}([\operatorname{AvgPool}(F) ; \operatorname{MaxPool}(F)])\right) \\
    & =\sigma\left(f^{7 \times 7}\left(\left[F_{\text {avg }}^s ; F_{\max }^s\right]\right)\right)
\label{eq2}
\end{aligned}
\end{equation}

    where $W_0$, $W_1$ denotes for MLP weight ($W_0 \epsilon R^{C/r \times C}$, $W_1 \epsilon R^{C \times C/r}$ with $r$ is reduction ratio), $\sigma$ symbolize for Sigmoid activation function and $F_{avg }^c, F_{\max }^c$ corresponding for generated feature maps from average and maximum pooling, respectively. The two attention modules will compute input feature maps sequentially, and then multiply these attention maps according to the input feature maps in order to transform the features according to their inputs. Based on Eq ~\ref{eq1} and Eq ~\ref{eq2}, calculation of the CBAM module is shown in Eq ~\ref{eq3} and the structure of CBAM block is illustrated in Fig ~\ref{fig1}.

\begin{equation}
\begin{aligned}
    F^{\prime} & =M_c(F) \otimes F \\
    F^{\prime \prime} & =M_s\left(F^{\prime}\right) \otimes F^{\prime} \\
    F_O & =I \oplus C B S\left(F^{\prime \prime}\right)
\label{eq3}
\end{aligned}
\end{equation}

\begin{figure}
    \centering
    \includegraphics[width=0.9\linewidth]{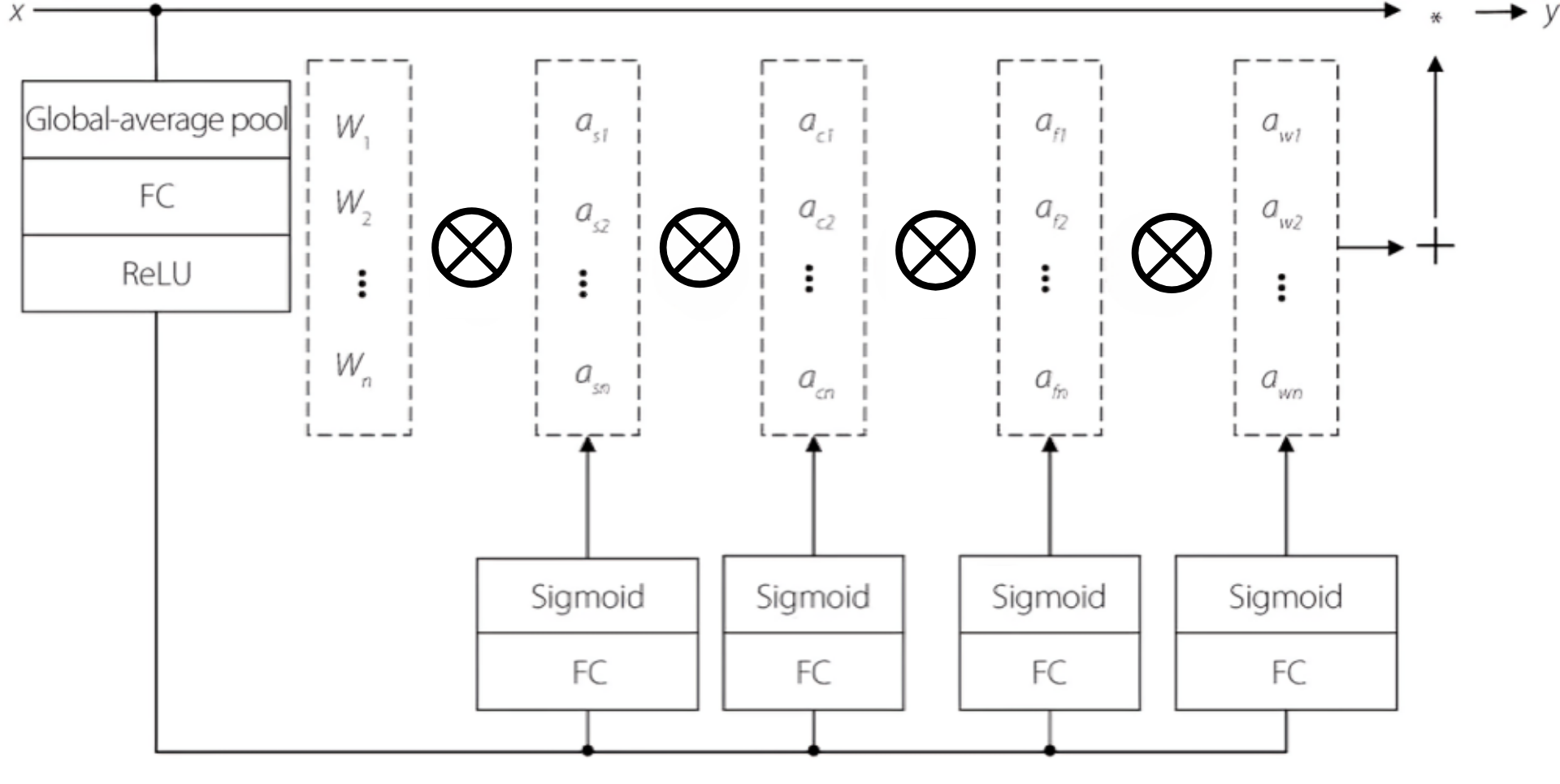}
    \caption{ODConv Detailed Structure}
    \label{fig2}
\end{figure}

    where $\otimes$, $\oplus$ denotes for element-wise multiplication and addition, respectively, and $F^{\prime}$ is the result form channel attention, $F^{\prime \prime}$ is the result form spatial attention and $F_O$ is the output of the CBAM block. In this paper, we replace the standard convolutions in spatial attention module of CBAM to ODConv. The ODConv is created base on DyConv \cite{b15} and CondConv \cite{b16}. Unlike traditional convolutions with fixed weights, ODConv generates kernel weights dynamically, allowing for more flexible and input-dependent feature extraction. This adaptive feature enables ODConv to capture more complex and varied patterns in the data as especially in MFCCs data use for this paper, the calculation of ODConv is outlined in Eq ~\ref{eq4} and Fig ~\ref{fig2} illustrated the structure of the ODConv with $W_i$ stand for convolution kernel, in the FC layer compressed feature vector is mapped into a low-dimensional space with a reduction ratio of $\gamma$ \cite{b17}.

\begin{equation}
\begin{aligned}
    y &= (\alpha_{w1} \otimes \alpha_{f1} \otimes \alpha_{c1} \otimes \alpha_{s1} \otimes w_1 + \cdots \\
      &\quad + \alpha_{nm} \otimes \alpha_{fm} \otimes \alpha_{cm} \otimes \alpha_{sm} \otimes w_m) * x
\label{eq4}
\end{aligned}
\end{equation}

    where $\alpha_{wi}$ denotes for scalar attention of the kernal of $W_i$ and $\alpha_{ci} \epsilon R^{Cin}$, $\alpha_{fi} \epsilon R^{Cout}$, $\alpha_{si} \epsilon R^{k \times k}$ denotes for three freshly new attentions, computed along the spatial dimension, output dimensions and input dimensions and $*$ symbolize for multiplication.

    The spatial attention module, which previously utilized a standard convolutional layer to produce a spatial attention map, now uses ODConv to generate spatial-wise attention weights. The integration of ODConv into CBAM is expected to enhance the module's ability to adapt to different input features, potentially leading to improved attention mechanisms. By dynamically adjusting the convolution kernels based on input characteristics, the modified CBAM can potentially capture more nuanced spatial and channel-wise dependencies.

\subsection{Dynamic-Attention Network}

    In this paper, we proposed an architecture for emotion classification from audio data from spectral features and furthur more we also tested many different structures with many processing stream and methods including dual-stream from both raw waveform and MFCCs. The first stream of the model, the input is sound waves with length of 5 seconds and a frequency of 16 kHz which have been vectorization to go through the embedding layers for feature extraction four layers of 1D CBM layers (Convolution-Batchnormalize-Maxpooling), this hierarchical structure allows the model to learn increasingly abstract representations of the temporal dynamics in the audio signal. Then, the output of these convolutional layers is then fed into a Bidirectional Gated Recurrent Unit (Bi-GRU) layer \cite{b18}, the Bi-GRU is capable of capturing long-term dependencies in both forward and backward directions of the audio sequence, which is particularly valuable for emotion recognition tasks where context is crucial. The calculation of GRU layer is shown follow by Eq ~\ref{eq5}.

\begin{equation}
\begin{aligned}
r_t &= \sigma(W_{ir} x_t + b_{ir} + W_{hr} h_{t-1} + b_{hr}) \\
z_t &= \sigma(W_{iz} x_t + b_{iz} + W_{hz} h_{t-1} + b_{hz}) \\
n_t &= \tanh(W_{in} x_t + b_{in} + r_t \ominus (W_{hn} h_{t-1} + b_{hn})) \\
h_t &= (1 - z_t) \ominus n_t + z_t \ominus h_{t-1}
\label{eq5}
\end{aligned}
\end{equation}

    where $x_t$ is the input of time $t$, $h_{(t - 1)}$ is the hidden state of the layer at time $t-1$ or the initial hidden state at time $o$, and $r_t, z_t, n_t$ are the reset, update, and new gates, respectively, $\ominus$ and is the Hadamard product, $h_t$ is the hidden state at time $t$. In Bi-GRU, the input $x_t^2$ for the current layer is the hidden state $h_t^1$ of the previous layer multiplied by dropout factor $\delta_t^1$ where each $\delta_t^1$ is a Bernoulli random variable, this means that the input to the current layer is a randomly dropped-out version of the hidden state from the previous layer, which can help prevent overfitting and improve the model's generalization performance. The second stream begins by transforming the raw audio into Mel-Frequency Cepstral Coefficients (MFCCs), the data will be exported and imported into the MFCC to learn the time sequence after four layers of 2D CBM layers. The output from embedding layers then passed through the proposed Dynamic-CBAM block, this will enhances the representational power of the network by adaptively refining feature maps along both channel and spatial dimensions, potentially highlighting the most emotionally salient aspects of the spectral representation.
    
\begin{figure}
    \centering
    \includegraphics[width=0.8\linewidth]{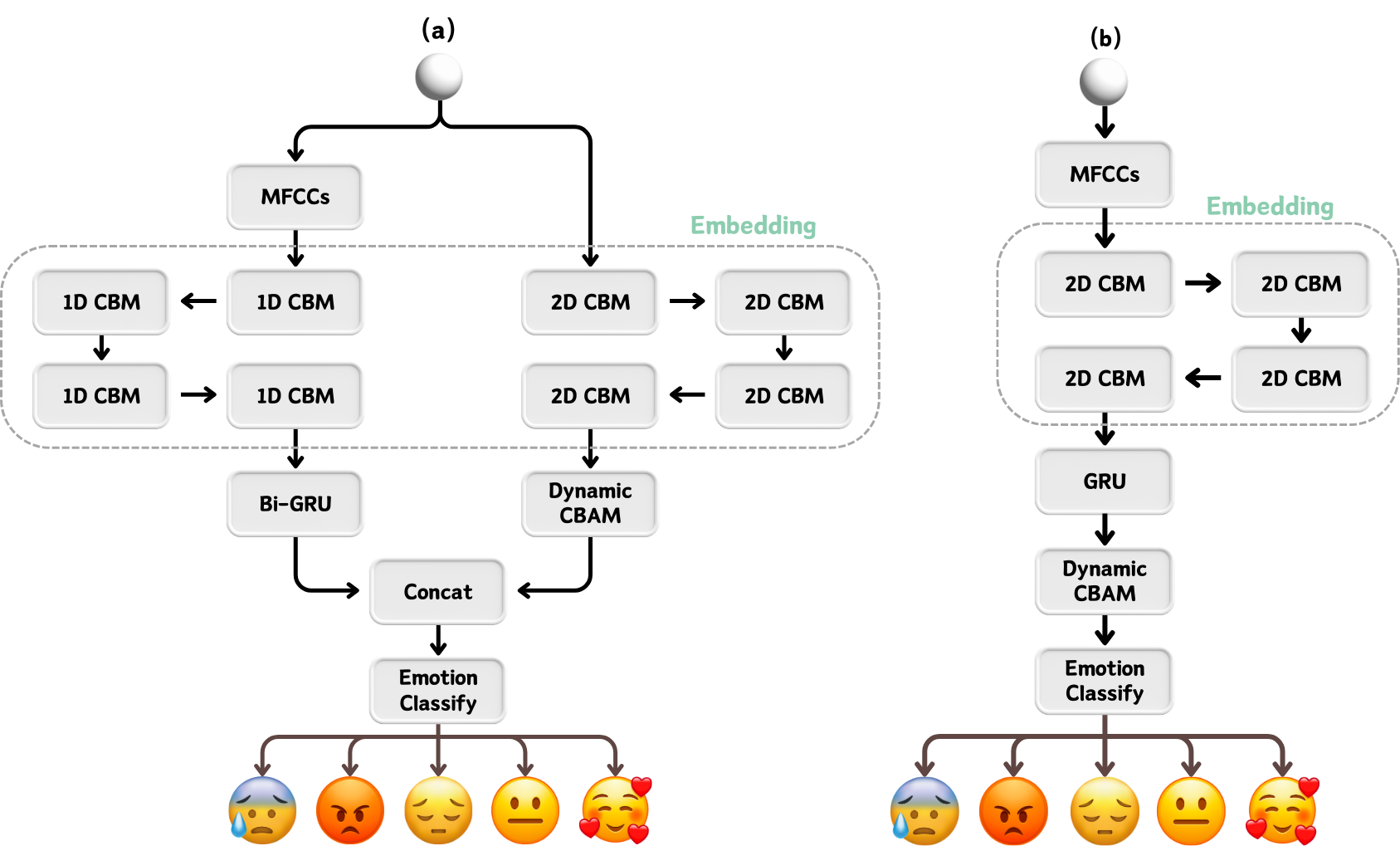}
    \caption{(a) Experiment Model Structure with Dual-stream (b) Proposed Dynamic Attention Network Architecture}
    \label{fig3}
\end{figure}

    After calculating through a series of cumulative layers, the data will pass through the concentrate to combining the complementary information extracted from the raw waveform and spectral features, and then the features will pass through a classification layer to produce the final output from 0 to 4, corresponding to the five types of emotions: anger, happiness, sadness, fear, and neutral. In this study, to test the effectiveness of the model, we have tested many different structures with many processing stream and methods. The architectural of model is outlined in Fig \ref{fig3}, with (a) having two streams, it's allows the model to leverage both the fine-grained temporal information from the raw waveform and the spectral characteristics captured by the MFCCs and in (b) is the proposed model with only stream by using only MFFCs to avoid complexity, this model does not use waveform data to avoid data corrupted spike-noises in the raw waveform data.

\section{Experiment}
\subsection{Dataset}

    In this research, we use the Vietnamese Speech Emotion dataset (VNEMOS) \cite{b19} that we have compiled in recent work, the dataset contains both natural and acted data. Natural data is gathered from conversations and live TV shows, while in the acting part, it is gathered from movies and live shows. We collected 250 segments with approximately 30 minutes from 27 movies, movie series and live shows. By dividing the data into five main basic emotions of humans “anger, happiness, sadness, neutral and fear”, we assure that our dataset provides an exhaustive and diverse spectrum of emotional states.

\subsection{Pre-processing}

\begin{figure}[htbp] 
    \centering
    \includegraphics[width=0.7\linewidth]{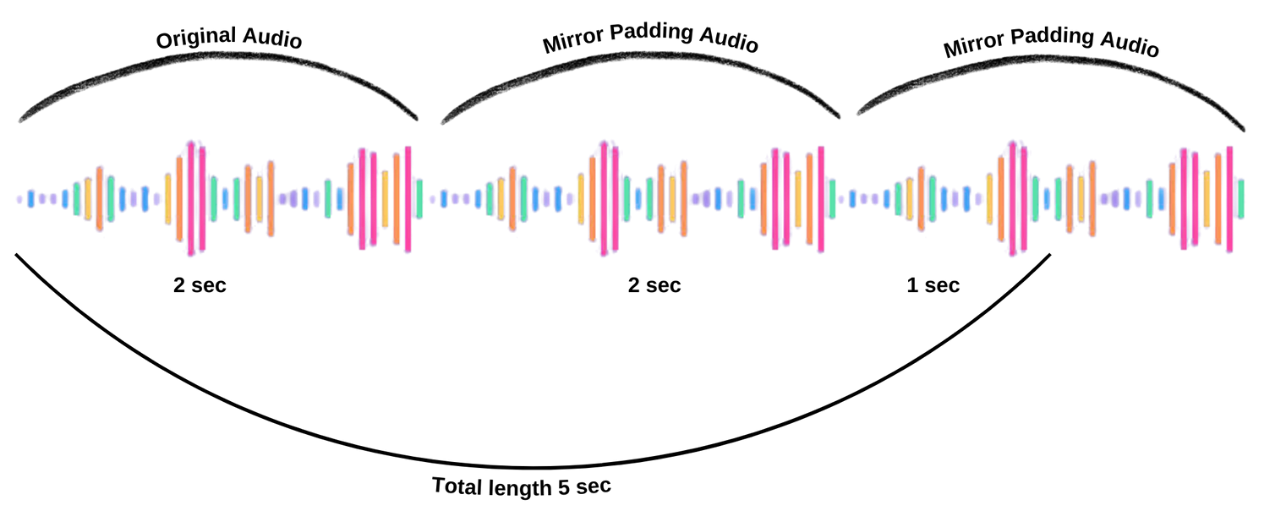}
    \caption{\textit{Mirror-padding Algorithm}}
\label{fig4}
\end{figure}

    Audio signal frequency content as it changes by dividing it into its constituent frequency components over a short period of time, which allows one to analyze the frequency content as it changes over time. In this process, audio data in one dimension got resample with 160kHz and segmented into windows with 8kHz for two-way MFCC parameterization group. To calculate these parameters, the speech signal is converted to spectrum, followed by fast Fourier Conversion and finally a cepstral coefficient is calculated. Inspired by symmetric padding used in image processing, we developed a method that uses audio-specific padding. Mirror padding is a technique that duplicates the sound waves in the original audio segment. By using this method, we are able to use audio clips that were of shorter length than the length necessary for our research due to this method of duplicating the original audio clip to the required length. As an example of how mirror-padding works, if we need a 5-sec audio segment but the data we have is only 2 seconds, the mirror padding job will be to make the audio segment as long as the requirements length. It will repeat the audio segment so that the audio segment is long enough compared to the specified length to help prevent data errors, mirror-padding algorithm is illustrated in Fig ~\ref{fig4}.

\subsection{Environment Setup}

    In this research, to ensure that the accuracy and reliability of the results, we utilized K-fold stratified cross-validation to ensure that the distribution of all classes in the training and test datasets is consistent with the class distribution in the original dataset. This approach maintains the class proportions in each fold, leading to a more accurate assessment of the model's performance across different classes with 5-Fold cross-validation. This method assesses the model by dividing the dataset into 5 equal parts, each part is used once as the test set while the remaining parts are used to for the training process, the results from these folds are then averaged to provide a final evaluation of the model performance. We have trained and evaluate the model with hyperparameters setting as $epoch = 100, lr = 0.001, batch_size = 32, optimizer = adam$ using environment as outlined in Table ~\ref{tab3}.

\begin{table}[h!]
    \centering
    \caption{Environment Resources Configuration.}
    \begin{tabular}{|l|l|}
        \hline
        \textbf{Component} & \textbf{Details} \\
        \hline
        CPU & Intel® Core™ i5-8400 \\
        \hline
        GPU & NVIDIA RTX GeForce 2080Ti \\
        \hline
        RAM & 32Gb \\
        \hline
        Python Version & Python 3.11.9 \\
        \hline
        Host Environment & Window 11 Pro 64 bits \\
        \hline
    \end{tabular}
\label{tab3}
\end{table}

\subsection{Loss Function}

    \normalsize{For our research, we use Cross-Entropy Loss as the loss function for training our models. Cross-Entropy Loss is widely used for classification problems and is particularly effective when dealing with multi-class classification tasks, the loss function is defined as in Eq ~\ref{eq6}.

\begin{equation}
    \text{ Cross-Entropy Loss} = - \sum_{i=1}^{N} y_a \log(\hat{y}_a)
\label{eq6}
\end{equation}

    where $N$, $y_a$ denotes for the number of classes and the binary indicator (0 or 1) if class label $a$ is the correct classification for a given observation, respectively, and $\hat{y}_i$ is the predicted probability of the observation belonging to class $a$.

\subsection{Evaluation Metrics}

    To quantitatively evaluate the performance of the proposed model, three metrics were used as evaluation criteria for this research, first "Precision" represents the ratio between the number of true positive predictions of the total number of predicted positive cases, precision measures the accuracy of positive predictions, "Recall" represents the ratio of true positive predictions to actual positive cases in the data set, it measures the classifier's ability to find all positive cases, and the "F1 Score" is the harmonic mean of precision and recall provides a balance between precision and recall. Weighted accuracy (WA) takes into account the class distribution when calculating the accuracy, this is particularly important when dealing with imbalanced datasets, where some classes are significantly more prevalent than others and Unweighted accuracy (UA) treats all classes equally, regardless of their frequency in the dataset, it's the simple average of the accuracies for each classes, by using UA and WA, the evaluation process provides a balanced perspective where UA highlights the model generalization ability across all classes, while WA indicates its effectiveness in handling the class distribution, the detailed evaluation equation used is shown in Eq ~\ref{eq7} - ~\ref{eq10}.
    
\begin{equation}
    \left\{\begin{aligned}
    \text{P} = \frac{TP}{TP + FP} \\
    \text {R} = \frac{TP}{TP+FN} \\
    \text {F1-score} = 2 \times \frac{{P} \times {R}}{{P}+{R}}
    \end{aligned}\right.
\label{eq7}
\end{equation}

\begin{equation}
    \text {UA} = \frac{TP + TN}{TP + FN + TN + FN}
\label{eq9}
\end{equation}

\begin{equation}
    WA = \frac{1}{n}\sum_{o=1}^{n}\frac{TP_o + TN_o}{TP_o + FN_o + TN_o + FN_o}
\label{eq10}
\end{equation}

    where $TP$, $TN$ represent the number of correct identifications and corrective false identifications according to the ground truth emotion labels, $FP$ indicates the number of incorrect identifications when it should have been positive and $FN$ indicates the number of genuine cases that have not been classify. With $o$ is the total number of classes in the dataset and metric with $o$ defined for it values in $o$-th class, respectively.

\section{Performance evaluation and Discussion}

    For this research we use vietnamese emotion dataset, which included a number of emotions such as anger, sadness, happiness, anxiety, and neutral for the depression diagnosis task. In this paper, we have evaluate the model with many difference setting to ensure the robustness of the proposed model and proposed methodology. Tables ~\ref{tab4} shown classifying results the five emotions using features derived form the VNEMOS dataset using MFCCs and waveform data is presented, respectively.

\begin{table}[h!]
\centering
\caption{Evaluation results of diference type for comparation with proposed model.}
\begin{tabular}{|l|l|c|c|c|}
\hline
    \textbf{Model} & \textbf{Data} & \textbf{UA } & \textbf{WA} & \textbf{F1-score}\\
    \hline
    One-stream & Soundwave & 0.73 & 0.72 & 0.73\\ 
    \hline
    One-stream GRU & Soundwave & 0.80 & 0.79 & 0.80 \\ 
    \hline
    One-stream GRU & MFCCs & 0.82 & 0.81 & 0.82 \\ 
    \hline
    One-stream Bi-GRU & Soundwave & 0.76 & 0.75 & 0.76 \\ 
    \hline
    Dual-stream Bi-GRU & Soundwave, MFCCs & 0.84 & 0.83 & 0.84\\ 
    \hline
    Dual-stream Dynamic-CBAM & Soundwave, MFCCs & 0.85 & 0.85 & 0.85 \\ 
    \hline
    Dual-stream Dynamic-CBAM Bi-GRU & Soundwave, MFCCs & 0.86 & 0.85 & 0.86 \\ 
    \hline
    \textbf{Proposed Model} & MFCCs & \textbf{0.87} & \textbf{0.86} & \textbf{0.87}\\ 
    \hline
    Anh, N. Q., et al. \cite{b19} & MFCCs & 0.85 & 0.83 & 0.85 \\ 
    \hline
\end{tabular}
\label{tab4}
\end{table}

    The evaluation results presented in Table ~\ref{tab4} shown a comprehensive comparison of various speech emotion recognition models with it input datas. The Proposed Model, utilizing only MFCCs as input data, achieves the highest performance across all metrics with 0.87 UA, 0.86 WA, 0.87 F1-score, slightly outperforming the Dual-stream Dynamic-CBAM model setting which uses both Soundwave and MFCCs. This suggests that the proposed model architecture effectively extracts emotion-relevant features from MFCCs alone, potentially offering computational efficiency advantages. The results from One-stream 0.73 UA to One-stream Bi-GRU 0.80 UA to Dual-stream Bi-GRU 0.84 UA have demonstrated the benefits of incorporating bidirectional GRU and dual-stream processing. Notably, the introduction of Dynamic-CBam in the dual-stream configuration with 0.86 UA yields a significant performance boost, highlighting the effectiveness of this attention mechanism in capturing relevant emotional cues. The Dual-stream Dynamic-CBam Bi-GRU model shows a slight performance decrease compared to its counterpart without Bi-GRU, suggesting that the additional complexity might not always translate to improved performance in this context. Finally, the proposed model have better performance compared to the Previous work with 0.85 UA, this underscores the advancements made in the current study. Overall, these results have illustrated the efficacy of dual-stream processing, the power of Dynamic-CBAM attention mechanisms and the potential of the proposed architectures to extract emotion-relevant information effectively from acoustic features. To further visualize the effectiveness and robustness of the proposed model, Fig ~\ref{fig5} have oulined the confusion matrix of the model on VNEMOS dataset.

\begin{figure}
    \centering
    \includegraphics[width=0.75\linewidth]{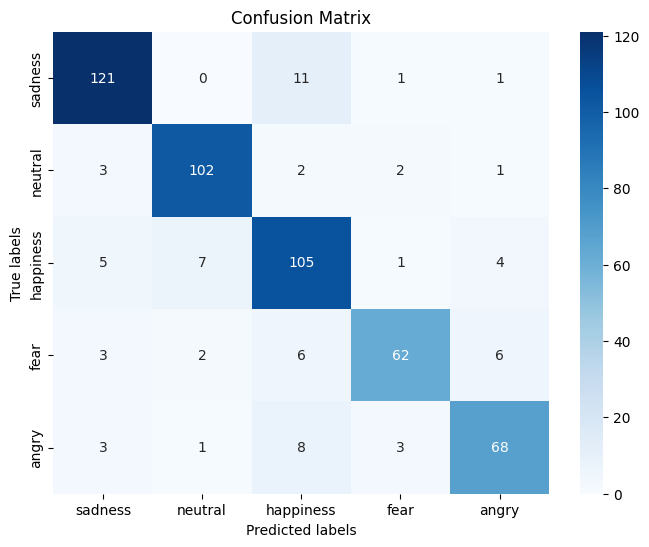}
    \caption{Confusion Matrix from the Proposed Model}
    \label{fig5}
\end{figure}

\section{Conclusion}

    In this study, the application of deep neural networks to identify emotions from speech to recognize depression tendency. In particular, a Dynamic Attention GRU Network have been proposed and aslo in this paper, we have introduce a new block named Dynamic-CBAM. All in all the research has an impressive UA, WA, F1 with 0.87, 0.86 and 0.87, respectively. This is the prerequisite step for developing the model after editing and changing the adjustments to be able to have higher rates and be able to differentiate in special situations. In the future we will try to improve the quality of the model and also improve the accuracy, also it's possible to identify audio segments that are free of noise or distortions and recognize emotional data without the hindrance of language barriers and can help improve sound quality.

\end{document}